\documentstyle[aps,pre]{revtex}
\begin{document}
\draft
\bibliographystyle{prsty}
\title{Universality and Scaling for the Structure Factor in 
Dynamic Order-Disorder Transitions}
\author{Gregory Brown$^{1,2}$, Per Arne Rikvold$^{1,2,3}$, 
and  Martin Grant$^2$}
\address{ $^1$  Center for Materials Research and Technology,\\
                Supercomputer Computations Research Institute,
                and Department of Physics\\
	        Florida State University, Tallahassee, Florida 32306-4350\\
          $^2$  Centre for the Physics of Materials, Physics Department, \\
                Rutherford Building, McGill University,\\
	        3600 rue University, Montr{\'{e}}al, 
		Qu{\'{e}}bec, Canada H3A 2T8\\
	  $^3$  Department of Fundamental Sciences,
	        Faculty of Integrated Human Studies\\
	        Kyoto University, Kyoto 606, Japan\\
}

\date{\today}
\maketitle
\begin{abstract}
The universal form for the average scattering intensity from systems
undergoing order-disorder transitions is found by numerical
integration of the Langevin dynamics. The result is nearly identical
for simulations involving two different forms of the local
contribution to the free energy, supporting the idea that the Model A
dynamical universality class includes a wide range of local
free-energy forms. An absolute comparison with no adjustable
parameters is made to the forms predicted by the theories of
Ohta-Jasnow-Kawasaki and Mazenko. The numerical results are well
described by the former theory, except in the cross-over region between
scattering dominated by domain geometry and scattering determined by
Porod's law.
\end{abstract}

\section{Introduction}

Phase ordering by quenching from a region of the phase diagram where a
material is uniform to one where several phases coexist at equilibrium
provides an important technique for creating multi-phase materials
that are inhomogeneous on mesoscopic length scales. Since the
macroscopic properties of such materials can be quite different from
those of the constituent phases and depend sensitively on the
mesoscopic structure, a solid understanding of pattern formation in
phase-ordering systems is important to several branches of materials
science. Examples include precipitation strengthening in
metals \cite{ZAND97} and fabrication of glasses\cite{TOMO86}.

Universality allows one to describe the structure and dynamical
properties of diverse phase-ordering systems using models that only
take into account properties such as conservation laws and
order-parameter symmetries. Two important universality classes involve only
local relaxational dynamics and an order parameter that can be
represented as a scalar field \cite{HOHE77,BRAY94}.  The order
parameter is not conserved in the first class, called Model A. This
can be used, for example, to model anisotropic magnets and alloys undergoing
order--disorder transitions.  In the second class, called
Model B, the order parameter is a locally conserved quantity, and
relaxation proceeds by diffusion away from regions of high chemical
potential. When hydrodynamic modes and strain effects can be ignored,
binary mixtures and alloys are described by this model.

Scaling is the hypothesis that the behavior of the system over a large
range of length scales can be described in terms of a single
characteristic length $R$. A necessary condition for scaling is that
$R$ must be well separated from any other microscopic or macroscopic
length scales present in the system. For many phase-ordering
processes, the characteristic length has been found to have a
power-law dependence on the time, $\tau$, elapsed since the quench,
$R\sim\tau^n$.  In non-conserved systems the power law is readily
observable with $n=1/2.$ For conserved systems the late-time growth
exponent is $n=1/3$, although processes occurring at early times may
mask this behavior. Both of these universality classes are
important. However, the work reported here is restricted to
non-conserved systems, for which more detailed theoretical
results are available.

Here, scaling and universality in phase ordering are tested by
comparing numerical solutions of the Langevin equation involving two
different forms for the local part of the free-energy functional: the
$\psi^4$ Ginzburg-Landau form and a piecewise-linear triangular
form. The numerical details are presented in Sec.~II. The agreement
between the structure factors found for both forms is discussed in
Sec.~III.  Combined with a mapping to sharpen the interfaces,
numerical solution of the Langevin equation allows comparison to
analytic theories for the structure factor, {\it without\/} fitting
parameters. In Sec.~IV three theories are reviewed and compared to the
numerical results. The Ohta-Jasnow-Kawasaki theory agrees with the
simulations at small and large wavevectors with noticeable deviations
only for intermediate wavevectors. The theories by Mazenko are
qualitatively correct, but the zeroth-order theory agrees with the
simulations better than the second-order theory. Sec.~V is a brief
summary of our results.

\section{Numerical Model}

A general model of phase ordering can be constructed from a free
energy composed of a local term with two degenerate minima and a
nonlocal term representing the contribution from spatial fluctuations
\cite{HOHE77},
\begin{equation}
\label{Potential}
{\cal F}[\psi({\bf r}, \tau)] 
= \int d{\bf r} \left\{f\left[\psi({\bf r},\tau)\right]
+\frac{1}{2}|{\bf\nabla}\psi({\bf r},\tau)|^2 \right\} \;,
\end{equation}
where the scalar order-parameter field $\psi({\bf r},\tau)$ depends on
position $\bf r$ and time $\tau$. The dynamics of the model are
governed by a Langevin equation,
\begin{equation}
\label{eq:Leq}
\frac{\partial \psi({\bf r},\tau)}{\partial \tau} = 
-\frac{\delta {\cal F}[\psi({\bf r},\tau)]}{\delta \psi({\bf r},\tau)}
+\sqrt{\epsilon}\eta({\bf r},\tau) \;,
\end{equation}
where the first term on the right-hand side corresponds to
deterministic relaxation towards a minimum of the free energy, and the
second term represents thermal fluctuations. The thermal noise is
assumed to be Gaussian with zero mean and correlations given by
\begin{equation}
\langle \eta({\bf r},\tau)\eta({\bf r}',\tau') \rangle =
\delta({\bf r}-{\bf r}')\delta(\tau-\tau') \;.
\label{eq:FDrel1} 
\end{equation}
The strength of the noise is given by the normalized temperature
$\epsilon$, which is the only parameter in the model after rescaling
the order parameter, space, and time\cite{StrengthNote}. The primary
effect of thermal fluctuations is to introduce randomness at early
times\cite{HERN92}.  In the late-time scaling regime the dynamics of
this model are controlled by a zero-temperature fixed point, and
thermal fluctuations can be ignored.

The local part of the free-energy functional $\cal F$ is usually
chosen to have the Ginzburg-Landau form,
\begin{equation}
f\left[\psi({\bf r},\tau)\right] = -\frac{1}{2}\psi^2({\bf r},\tau) + 
\frac{1}{4}\psi^4({\bf r},\tau) \;,
\label{eq:f}
\end{equation}
which is a double well with degenerate minima. The corresponding
dynamical equation is
\begin{equation}
\frac{\partial\psi({\bf r},\tau)}{\partial \tau}
  = \left( 1 + \nabla^2 \right) \psi({\bf r},\tau) - \psi^3({\bf r},\tau)
  + \sqrt{\epsilon}\eta({\bf r},\tau) \;,
\label{eq:EOM}
\end{equation}
which is commonly called the time-dependent Ginzburg-Landau (TDGL)
equation.

To test universality for non-conserved scalar order-parameter systems,
comparisons need to be made with a model different from the standard
TDGL model. The new model still needs to follow the non-conserved
dynamics of Eq.~(\ref{eq:Leq}) for a scalar order parameter with two
degenerate equilibrium values. One obvious aspect to change is the
form of the local part of the free-energy functional, 
$f[\psi({\bf r},\tau)]$. A choice quite different from the 
Ginzburg-Landau form is a piecewise-linear triangular double well,
which is not harmonic near the minima. This form is compared to the
$\psi^4$ Ginzburg-Landau potential in Fig.~1. It has been chosen to
match the Ginzburg-Landau potential at the extrema, {\it i.e.\/}
$f(\pm 1)=-1/4$ and $f(0)=0$. The Langevin equation for this potential
is
\begin{equation}
  \frac{\partial \psi({\bf r},\tau)}{\partial \tau}
  = \sqrt{\epsilon} \eta({\bf r},\tau) + \nabla^2 \psi({\bf r},\tau) - 
  \left\{ \begin{array}{ll}
  -1/4, & \ \ \psi < -1 \\
  +1/4, & \ \ -1 \le \psi < 0 \\
  -1/4, & \ \ 0 \le \psi < +1 \\
  +1/4, & \ \ \psi \ge +1 
\end{array} \right. \;. 
\label{eq:triEOM} 
\end{equation}

The simulations with both potentials were conducted on square lattices
with periodic boundary conditions and lattice constant $\Delta
r=1\,$. The system size considered was $L_x=L_y=L=1024$.  The
Laplacian was approximated by an $8$-point form \cite{BROW97a}, and
and a simple Euler integration scheme with $\Delta \tau=0.05$ was used to
collect data every $25$ time units up to a maximum of $\tau=2000$.
Results for the TDGL equation were averaged over $100$ realizations;
only $10$ realizations of the piecewise-linear model were needed to
confirm its agreement with the TDGL results.

\section{Universality}

The average structure of the system at a specific time after the
quench can be described by the order-parameter correlation function
\begin{equation}
C(|{\bf r}-{\bf r}'|,\tau) = 
  \langle \psi({\bf r},\tau) \psi({\bf r}',\tau) \rangle \;,
\end{equation}
where $\langle \cdots \rangle$ denotes averaging over the ensemble of
quenches as well as the volume. When scaling holds, the correlation
function can be expressed in terms of a time-independent master curve,
\begin{equation}
C(r,\tau)=\bar{C}\left(\bar{r}\right) \;.
\end{equation}
where $\bar{r}=r/R(\tau)$.  While microscopy techniques can be used to
measure $C$, scattering experiments probe its Fourier transform,
$S({\bf k},\tau)$, called the structure factor. The structure factor
is also related to a time-independent form,
\begin{equation}
S\left(k,\tau\right) = R^d(\tau) F(q)
\end{equation}
where $q=kR(\tau)$ and $d$ is the dimension of the system.

When infinitely sharp domain interfaces are randomly oriented
throughout the system, the small-${\bar r}$ form of the correlation
function is linear, $\bar{C}(\bar{r})=1-\alpha\bar{r} + \cdots$, 
where $\alpha$
is proportional to the surface area to volume ratio for the domains
\cite{GUIN55}. This corresponds to a power-law decay of the scaled
structure factor at large $q$, $F(q)=(2\pi/\alpha)q^{-(d+1)}$, called
Porod's law \cite{BRAY94,GUIN55}. These results are valid for $1/q$
much smaller than the characteristic length, but much larger than the
domain interface width.

The assumption of infinitely sharp interfaces is used in most
situations, {\it e.g.} experimental and analytic studies. By design,
the domain interface in the simulations has a width of about
$\sqrt{2}$. We compensate for this by using a non-linear
mapping\cite{NONLIN}
of $\psi({\bf r},\tau)$ to $\pm 1$ before finding the Fourier
transform $\hat\psi({\bf k},\tau)$ \cite{BROW97a}. The structure
factor is
\begin{equation}
S({\bf k},\tau)=\langle|\hat\psi({\bf k},\tau)|^2\rangle \;,
\label{eq:SF}
\end{equation}
where $|\hat\psi({\bf k},\tau)|^2$ is the scattering intensity
associated with a particular domain pattern $\psi({\bf r},\tau)$, and
$\langle \cdots \rangle$ represents averaging over the ensemble of
quenches. The simulation results used to estimate the structure factor
were found by ``onion-shell'' binning of the two-dimensional
scattering intensity into a one-dimensional function\cite{BROW97a} of
$k$, averaging over all trials, and then scaling the result using the
average wavevector of each bin.

Including noise in the simulation has three drawbacks. The first is
purely practical: generating Gaussian random numbers for $\eta$ is
computationally expensive. The second drawback has a physical
basis. Ohta \cite{OHTA84} has shown that thermal fluctuations in the
TDGL equation retard growth of the characteristic length and cause a
broadening of $F(q).$ Thirdly, the thermal noise causes the structure
factor to cross over to the equilibrium $q^{-2}$ behavior for large
$q$, thus masking the $q^{-(d+1)}$ Porod tail. For these reasons, we
have simulated the TDGL equation at zero temperature with the
early-time fluctuations implemented by an initial condition where
$\psi({\bf r},\tau \! = \! 0)$ consists of independent random numbers
distributed uniformly on $[-0.1,+0.1]$.  

On the other hand, thermal fluctuations are essential for simulating
the piecewise-linear model. For this model at $T=0$, the domain growth
stops after a finite time. The time until the evolution stops, as well
as the corresponding characteristic length, depends on the random
initial condition. If a system which has stopped coarsening is briefly
heated, the growth continues for a while and then stops again. In
addition, making the potential shallower prolongs growth and yields
larger domains. The arrested growth appears to result from trapping
into metastable configurations associated with finite wavelength
corrugations on the interfaces. A constant, moderate level of thermal
noise completely prevents this trapping phenomenon, so noise with
$\epsilon=0.1$ was included in simulations of the piecewise-linear
model.

An estimate of $F(q)$ was constructed for each local form by combining
the estimates at different simulation times with the restrictions
$\tau>500$ and $k<0.5$. For $k$ larger than this, lattice effects
become important. The estimates of $F(q)$ for both models are
presented in Fig.~2, with $R(\tau)=\sqrt{2\tau}$ (see Sec. IV). The
agreement seen in Fig.~2(a), a log-log plot of $F$ vs.\ $q$, is very
good.  The Porod tail at large $q$ can be highlighted by plotting
$q^{d+1}F(q)$ against $q$, as shown in Fig.~2(b). This was found
independently of the structure factor, using $q=kR(\tau)$ before the
binning and averaging. The agreement is quite good well into the
power-law tail.  The claim that the details of the local part of the
free-energy functional do not influence the universal behavior appears
well supported by the numerical models.

\section{Analytic Theories}

Several theories for the universal form of the scaling function for
non-conserved systems exist. They can be compared to the numerical
integration of the TDGL equation without adjustable parameters.

One theory was developed by Ohta, Jasnow, and Kawasaki (OJK)
\cite{OHTA84,OHTA82}. The OJK theory starts from a diffusion equation
for a Gaussian auxiliary field, $u({\bf r},\tau)$,
\begin{equation}
\label{eq:DOJK}
\frac{\partial u({\bf r},\tau)}{\partial \tau} 
= D \nabla^2 u({\bf r},\tau),
\end{equation}
where the interfaces in the inhomogeneous material are defined to be
the set of ${\bf r}$ such that $u({\bf r},\tau)=0$, and the
order-parameter field is obtained by the mapping
$\psi({\bf r},\tau)={\rm sign}\left[u({\bf r},\tau)\right]$. The
coefficient $D=4\rho_d$ is the diffusion constant for the
interface. The factor $\rho_d=(d-1)/d$ results from assuming that the
interfaces are randomly oriented. As time progresses, the
characteristic size of the single-phase domains grows as $R_{\rm
OJK}(\tau)=\sqrt{D\tau}$, in accordance with the Lifshitz-Allen-Cahn
theory of domain growth driven by surface tension
\cite{LIFS62,ALLE79}. The two-point correlation function for the
order-parameter field in this model has the simple analytic form
\begin{equation}
\label{eq:COJK}
C_{\rm OJK}(r,\tau)=\frac{2}{\pi}
{\rm arcsin}\left[\exp{\left(-\frac{r^2}{2R_{\rm OJK}^2(\tau)}\right)}\right] 
\;.
\end{equation}
Since the system is isotropic, the Fourier transform of this correlation
function can be written in terms of a radial integral over a Bessel
function, and the scaling form of the structure factor is \cite{OHTA82,Rem1} 
\begin{equation}
F_{\rm OJK}(q)=\frac{2}{\pi}\frac{(2\pi)^{d/2}}{q}
\int_0^\infty {\rm d}w w^d
[\exp{(w^2)}-1]^{-1/2}(q w)^{1-d/2}J_{d/2}(q w) \;.
\label{eq:IOJK}
\end{equation}
Except for the dimensional dependence of $\rho_d$ in $R(\tau)$, this
result agrees with the perturbation calculation by Kawasaki, Yalabik,
and Gunton \cite{KAWA78}.

Another theoretical approach to phase ordering has been developed by
Mazenko \cite{MAZE89,MAZE90,LIU92,MAZE94b}. This theory also starts
with a Gaussian auxiliary field $u({\bf r},\tau)$. In contrast to the
OJK theory, $u({\bf r},\tau)$ is interpreted as the distance from
${\bf r}$ to the closest interface, and it is mapped onto the
order-parameter field using the equilibrium interface profile
$\psi({\bf r},\tau)=\psi_{\rm eq}\left(u({\bf r},\tau)\right)$. The
characteristic length is then defined in terms of the RMS distance
to the nearest interface, $R_0(\tau)=\sqrt{\pi\langle u({\bf
r},\tau)^2\rangle}$. Here the subscript $0$ is used to denote the
zeroth-order theory developed in
Refs.~\cite{MAZE89,MAZE90,LIU92}. Using the Gaussian properties of
$u({\bf r},\tau)$, the TDGL equation can be used to find a closed,
nonlinear partial differential equation which describes the scaling
form of the correlation function
\begin{equation}
\label{MazenkoEq}
 \frac{{\rm d}^2\bar{C}_0(v)}{{\rm d}v^2} 
+\left(\frac{d-1}{v}+\mu_0 v\right)
 \frac{{\rm d}\bar{C}_0(v)}{{\rm d}v}
+\tan{\left[\frac{\pi}{2}\bar{C}_0(v)\right]}=0 \;,
\end{equation}
where $v=r/R_0(\tau)$ is the scaled length. The characteristic length
is $R_0(\tau)=\sqrt{4\mu_0\tau}$ with $\mu_0$ an undetermined
parameter. The correlation function must obey the boundary condition
$\bar{C}_0(0)=1$. As a consequence, the third term in
Eq.~(\ref{MazenkoEq}) acquires a $1/v$ singularity at $v$=0. This
singularity must cancel against the $(d-1)/v$ factor in the second
term, resulting in the requirement that ${\rm d} \bar{C}_0(v)/{\rm d}
v = \sqrt{2/\pi (d-1)}$ at $v=0$.  Therefore, the first derivative of
$\bar{C}_0(v)$ at $v$=0 is not available to be used as the second,
independent constant in the solution of Eq.~(\ref{MazenkoEq}).
Instead, one must determine the value of $\mu_0$ such that
$\bar{C}_0(v)$ becomes integrable as $v \rightarrow \infty$. For $d=1$
the asymptotic form agrees with that found analytically for the
kinetic Ising model with Glauber dynamics \cite{GLAU63,AMAR90}, while
in the limit $d \rightarrow \infty$ the function $\bar{C}_0$ becomes
identical with the OJK result\cite{LIU92}. For finite $d>1$,
determining $\bar{C}_0$ is equivalent to the numerical solution of a
nonlinear eigenvalue problem \cite{MAZE90,Rem2,NumericalRecipes}.

Recently, Bray and coworkers \cite{BRAY94,BLUN93} and Yeung et
al.\ \cite{YEUN94}, have criticized the use of Gaussian fields in
constructing theories of phase ordering. Noting this criticism,
Mazenko \cite{MAZE94b} has expanded his theory to include a $u({\bf
r},\tau)$ whose distribution is an expansion around a Gaussian. The
theory with the second-order correction, here marked with the
subscript $2$, produces a new differential equation for the scaled
correlation function, $\bar{C}_2(v)$,
\begin{equation}
 \frac{{\rm d}^2 \bar{C}_2(v)}{{\rm d}v^2}
+\left(\mu_2 v+\frac{d-1}{v}\right)\frac{{\rm d}\bar{C}_2(v)}{{\rm d}v}
+\tan{\left\{\frac{\pi}{2}\left[\bar{C}_2(v)+\bar{H}_2(v)\right]\right\}}
-\frac{\pi}{2}
 \frac{\bar{H}_2(v)}{\cos^2{\left\{\frac{\pi}{2}\left[\bar{C}_2(v)
 +\bar{H}_2(v)\right]\right\}}}
= 0 \;.
\end{equation}
Here $\bar{H}_2(v)$ is a new function governed by
\begin{equation}
\frac{{\rm d}\bar{H}_2}{{\rm d}v}=\frac{b\bar{C}_2' - 
         \sqrt{(\bar{C}_2')^2+\alpha_0q_2b(1-b)}}
        {1-b}
\end{equation}
where $\bar{C}_2'={\rm d}\bar{C}_2/{\rm d}v$ and
\begin{equation}
b=-\frac{\pi}{2}\bar{H}_2(v)
  \tan{\left\{\frac{\pi}{2}\left[\bar{C}_2(v)+\bar{H}_2(v)\right]\right\}}
\;.
\end{equation}
A corrected eigenvalue $\mu_2 \ne \mu_0$ results, and the
characteristic length becomes $R_2(\tau)=\sqrt{4\mu_2\tau}$
with the scaled length $v=r/R_2(\tau)$. A new eigenvalue $q_2$ is
introduced, and the resulting double eigenvalue problem must solved
numerically\cite{Rem3}.

While experimental tests ultimately determine the validity of any
theory, finding the characteristic length {\it a priori\/} is difficult
for experiments, as well as many simulation techniques. The
characteristic length may be taken as a fitting parameter, with which
data can be fit well to all three theories. Such fitted comparisons
for Monte Carlo simulations made by OJK \cite{OHTA82} showed
quantitative disagreements in the tail of the correlation function.
Later Oono and Puri \cite{OONO88} used cell dynamics simulations to
show that the deviations might be caused by the nonzero interface
width present in simulations.  Blundell, Bray, and Sattler
\cite{BLUN93} also used cell dynamics simulations to test the
theories. To compare their simulation results to theory without
fitting, they reduced $R(\tau)$ to a parametric variable by plotting
values for the correlation function for the order parameter against
values for the correlation function for the square of the order
parameter for many $(r,\tau)$. The simulation results plotted this way
show good scaling, but the two Gaussian theories give the same result
for this particular scaling function \cite{BLUN93,YEUN94}.  As a
consequence, this method does not provide a strong test of the two
forms, which are, in fact, noticeably
different, as we will see below.  The question of how
well these theories predict the scaling function $\bar{C}(\bar{r})$
therefore remains open.

Numerical simulation of the TDGL equation can be compared to the theories
without fitting parameters since analytic forms for $R(t)$ are
known. We have chosen to work in terms of $R_{\rm OJK}$ since $R_0$
and $R_2$ depend on numerically determined eigenvalues.  The TDGL
estimates for the two-dimensional $F(q)$ for several times are
presented in Fig.~3, along with the predictions of the
theories. Fig.~3(a) is a log-log plot, showing good collapse of the
simulation data.  Concerning the difference 
between the theories of OJK and Kawasaki et al.\ \cite{KAWA78}, 
much better agreement with the simulations is
found using the factor $\rho_d$ predicted by OJK, so we
only consider that theory.  In Fig.~3(a) the
obvious differences between the theories shown are the magnitude at
small $q$ and the behavior near the shoulder at intermediate $q$. For
both of these features, the simulation data agree significantly better
with the OJK theory than with those of Mazenko.

Fig.~3(b) highlights the power-law tail at large $q$. The scaling of
the simulation data is quite good through the large peak for all of
the times presented here. Where the simulation data scales, it agrees
very well with the OJK theory.  There is a noticeable second peak in
the OJK and second-order Mazenko theories that is absent in the
zeroth-order Mazenko theory. Unfortunately, the simulation data cannot
be used to test the presence of a second peak because it does not
scale in this region for the practically attainable simulation
times. Instead, a pronounced trough is seen, which decays with
time. Simulations for significantly larger systems and longer times
would be needed to achieve asymptotic scaling in this region.

At values of $q$ higher than this, the scaling regime also has not
been reached, preventing an estimate of the Porod amplitude
($2\pi/\alpha$) for the simulations. However, a clear trend in the
data is apparent, indicating an amplitude which is significantly higher
than those predicted by both of Mazenko's theories. In fact, the
second-order theory has the smaller Porod amplitude. This is a result
of the relationship\cite{MAZE94b}
\begin{equation}
\label{eq:PorodAmp2}
\alpha_2=\alpha_0 
                \frac{\sqrt{q_2(q_2+2)}}{q_2+1} \;,
\end{equation}
where $q_2$ is the second-order eigenvalue. Since $\alpha_2>\alpha_0$
for any value of $q_2$, the predicted 
Porod amplitude is necessarily smaller for
the second-order theory. The disagreement with simulations increasing
with the order of the theory is also true for the decay of the
order-parameter autocorrelation function, although estimates for both
of Mazenko's theories are better than that for
OJK\cite{BROW97a,MAZE94b,LIU91b}.

Very recently, Emmott has presented a perturbation expansion about the 
OJK form for the correlation function, 
with a lowest-order term of order $1/d^2$ \cite{EMMO98}. 
Our numerical results presented here suggest that the OJK theory 
may provide a good starting point for a more complete theory of the 
dynamics of phase separation in systems with nonconserved order 
parameter. However, we note that the lowest-order correction 
provided by Emmott's approach would produce a change in
the Porod amplitude (which is well described by OJK), 
while it shares with OJK the inability to correctly describe the 
decay of the order-parameter autocorrelation function 
\cite{BROW97a,MAZE94b,LIU91b}.

\section{Conclusions}

Numerical simulations of the TDGL equation have been used to confirm
the universality of Model A systems with different forms for the local
free-energy contribution. With no adjustable parameters, the scaling
form of the structure factor from simulations is found to agree
quantitatively with the OJK theory at scaled wavevectors $q<3$. At
large $q$, the numerical results have not yet entered the asymptotic
scaling regime, even at our latest simulation times. However, the
trend is toward OJK and away from both the Mazenko theories. Although
the OJK theory has been found here to describe well the scaling form
of the structure factor, it does not correctly describe the decay of
the order-parameter autocorrelation function \cite{BROW97a,LIU91b}.

Mazenko's theories describe the estimated scaling form only
qualitatively. Interestingly, the zeroth-order theory gives better
agreement with simulations than does the second-order theory. Since,
aside from $d=1$, $d=2$ is expected to be the dimension where the
theoretical results are the most different, computational cost has
lead us to forgo three dimensional simulations at this time.  Another
aspect of universality not considered here is the effect of anisotropy
\cite{RUTE96}.

\section*{Acknowledgments}

We would like to thank G.~Mazenko for providing us with a numerical
determination of the second-order correlation function.  P.~A.~R.\ is
grateful for a useful conversation with K. Kawasaki and for
hospitality and support at McGill and Kyoto Universities. This
research was supported by the Florida State University (FSU) Center
for Materials Research and Technology, the FSU Supercomputer
Computations Research Institute under U.S.\ Department of Energy
Contract No.\ DE-FC05-85ER25000, and by U.S.\ National Science
Foundation grants No.\ DMR-9315969, DMR-9634873, and
INT-9512679. Research at McGill was supported by the Natural Sciences
and Engineering Research Council of Canada, and {\it le Fonds pour la
Formation de Chercheurs et l'Aide \`a la Recherche de Qu\'ebec\/}.
Supercomputer time at the U.~S.\ National Energy Research
Supercomputer Center was made available by the U.~S.\ Department of
Energy.


\begin{thebibliography}{10}
\bibitem{StrengthNote}
In the parameters and rescaling of Ref.~[7], $\epsilon=2k_BT
uc^{-d/2}a^{(d-4)/2}$. Using these parameters, the interface diffusion
constant of Eq.~(\protect{\ref{eq:DOJK}}) is $D=4\rho_d c\Gamma$.
\bibitem{Rem1} In our numerical calculations of the scaling functions,
we directly used the $d=2$ Fourier transform of Eq.~(\ref{eq:COJK})
without the integration by parts which leads to Eq.~(\ref{eq:IOJK}).
\bibitem{Rem2}
We constructed an implementation of this theory where values of $\mu_0$ were
  varied interactively. To guard against round-off errors, the eigenvalue
  problem was solved twice in 64-bit precision, once with a step size of
  10$^{-4}$ and once with 10$^{-3}$. The solutions agreed to six significant
  digits for $v \le 5.8$, beyond which point $\bar{C}_0 (v) < 10^{-8}$. We find
  $\mu_0 = 1.10420873646(1)$. The scattering intensity was constructed using a
  Romberg integration scheme and cubic spline interpolation from
  Ref.~\protect{\cite{NumericalRecipes}}.
\end{thebibliography}

\newpage
~
\begin{figure}[tbp]
\vskip 0.5in
\vskip 2.85in
\includegraphics{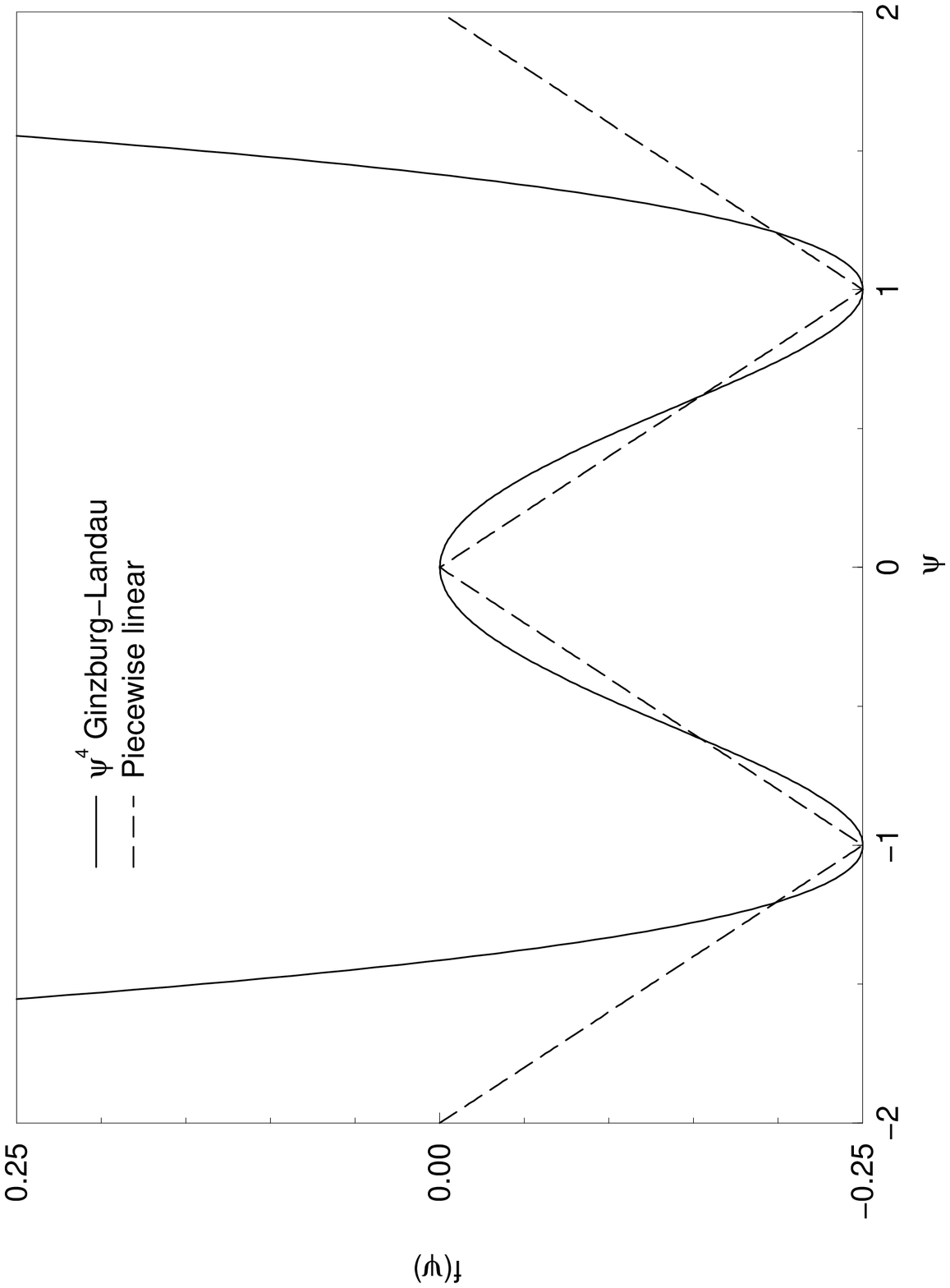}
\bigskip
\caption[]{Comparison of the two local parts of the free-energy
functional, $f[\psi({\bf r},\tau)]$, used in this study: the
Ginzburg-Landau form (solid curve) and a piecewise linear function
(dashed lines). Both potentials give the same estimate for the
structure factor, indicating that the Model A universality class does
not depend on the specific form of $f$.}
\end{figure}
\vfill
\newpage
~
\begin{figure}[tbp]
\vskip 3in
\includegraphics{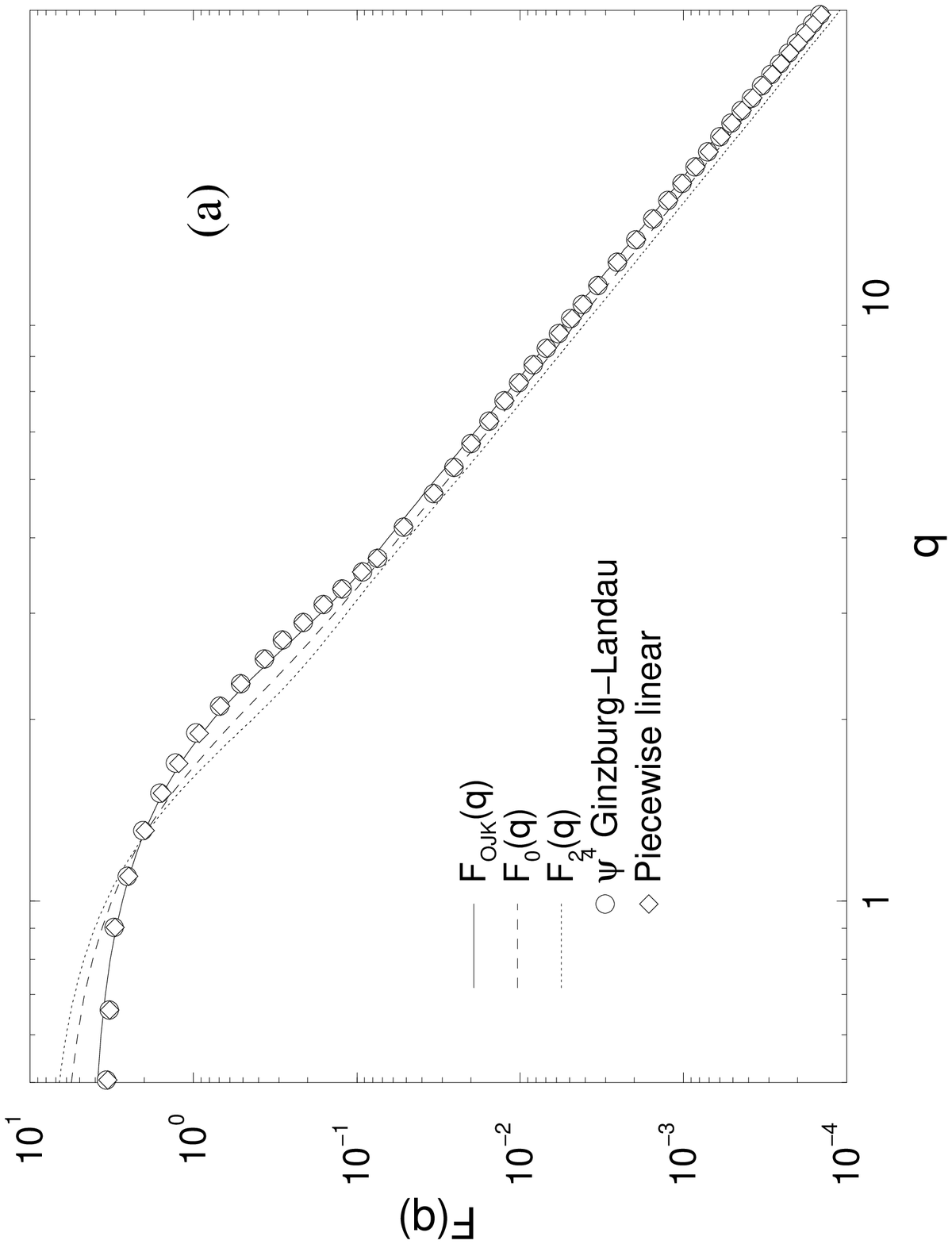}
\bigskip
\bigskip
\vskip 3in
\includegraphics{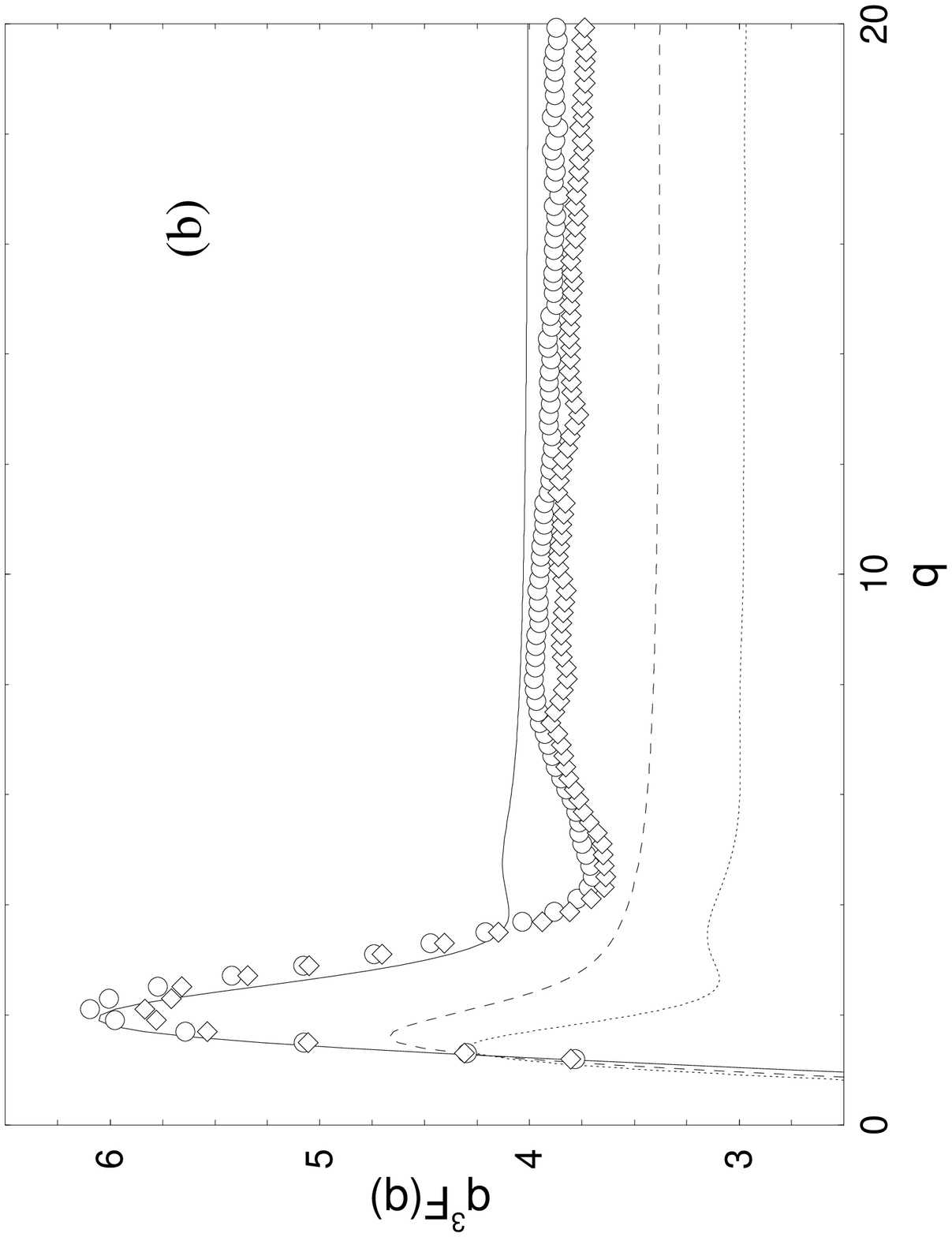}
\bigskip
\caption[]{Simulation estimates for both the Ginzburg-Landau ({\Large
$\circ$}) and piecewise-linear ($\diamondsuit$) models of (a) the
universal form of the structure factor and (b) scaling plot
emphasizing the Porod's law tail at large $q$. The estimates
incorporate all data for wavevectors $k<0.5$ and time $\tau>500$. The
scaling forms for the two models agree with each other, supporting the
idea that dynamical universality is not affected by the detailed shape
of the local part of the free-energy density $f[\psi]$. The OJK (solid
curve), Mazenko zeroth-order (dashed curve), and Mazenko second-order
(dotted curve) theories are included for reference. The simulations
show better agreement with the OJK theory than with the other
theories.}
\end{figure}
\vfill
\newpage
~
\begin{figure}[b]
\vskip 3in
\includegraphics{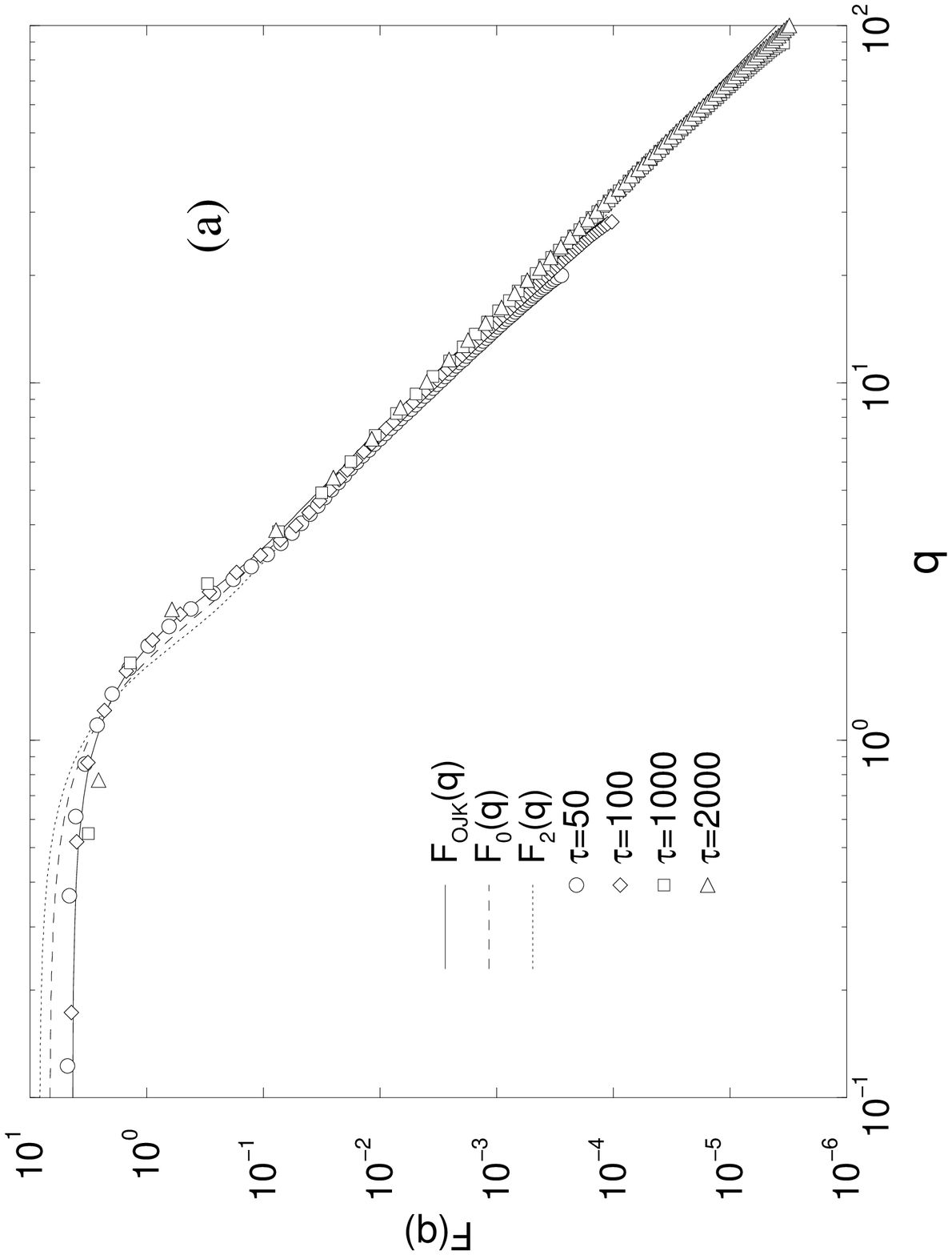}
\bigskip
\vskip 3in
\includegraphics{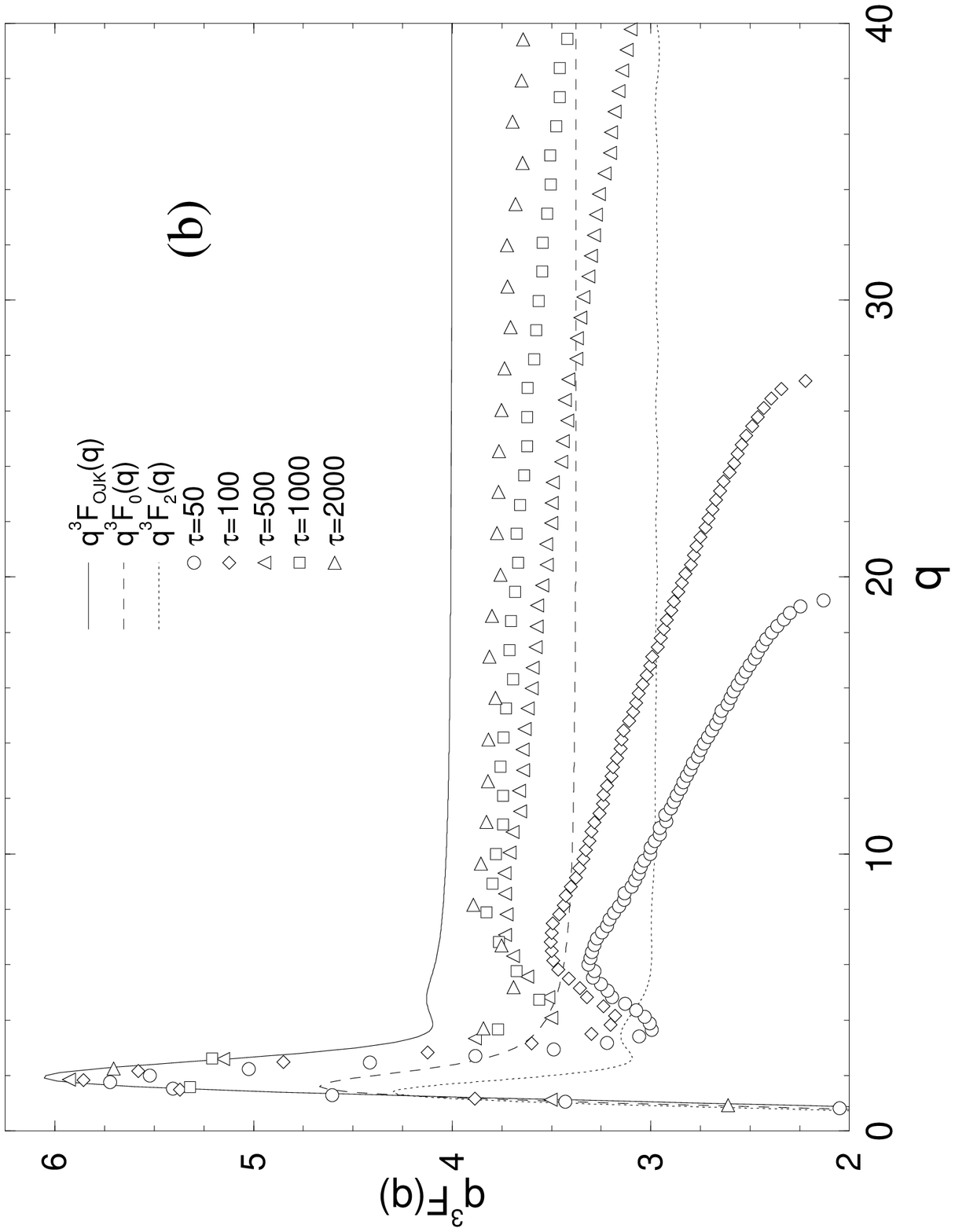}
\bigskip
\bigskip
\caption[]{Scaling of the structure factor for the two-dimensional
time-dependent Ginzburg-Landau model. The scaling variable is
$q=kR_{\rm OJK}(\tau)$. The simulation data for different times scale
quite well, using the characteristic length 
$R_{\rm OJK}(\tau)=\sqrt{4\rho_d\tau},$ 
except at large wavevectors where lattice effects are important. The
curves represent the theoretical scaling forms discussed in the text:
OJK (solid), Mazenko zeroth-order (dashed), and Mazenko second-order
(dotted). (a) For small values of $q$ ($q<3$), the simulations agree
quantitatively with the OJK theory. (b) Plot emphasizing the Porod's
law tail at large $q$, the trend with time indicates that the OJK
theory gives a better estimate of this amplitude than either of the
Mazenko theories. The theoretical amplitude of the second-order theory
is lower than that for the first-order theory.}
\end{figure}
\vfill

\end{document}